\documentclass{aa}  

\usepackage{graphicx}
\usepackage{txfonts}
\usepackage{graphicx}	
\usepackage{amsmath}	
\usepackage[T1]{fontenc}
\usepackage{lipsum}  
\usepackage{blindtext}
\usepackage{newtxtext,newtxmath}
\usepackage[normalem]{ulem}
\usepackage{hyperref}
\usepackage[switch]{lineno}

\hypersetup{colorlinks = true, linkcolor = {blue}, citecolor = {blue}, urlcolor= {blue}}

\newcommand{\simname}[1]{\texttt{#1}}
\begin{document} 

\title{ Dust-driven vortex cascades originating at water snow regions: A pathway to planetesimal formation}

   \titlerunning{Planetesimals in vortices at water snow regions} 
   \subtitle{}

   \author{Kundan Kadam\inst{1}
          \and
          Zsolt Regály\inst{2,3} 
          }

   \institute{Space Research Institute, Austrian Academy of Sciences, Schmiedlstrasse 6, A-8042, Graz, Austria\\
             \email{kundan.kadam@oeaw.ac.at}
         \and
             Konkoly Observatory, HUN-REN, Research Centre for Astronomy and Earth Sciences, Konkoly-Thege Mikl\'os 15-17, H 1121, Budapest, Hungary\\
            \email{regaly@konkoly.hu}
        \and
             ELTE Eötvös Loránd University, Institute of Physics and Astronomy, Department of Astronomy, Pázmány Péter sétány 1/A, H-1117, Budapest, Hungary.
             }

   \date{Received \today{}}

 
  \abstract
  {
  The origin of observed planetary systems, including our Solar System, as well as their diversity, is still an open question. 
Streaming instability (SI) is an important mechanism for the formation of gravitationally bound planetesimals, which can grow to form planetary embryos and eventually planets.
Snow lines in a protoplanetary disk can assist this process, as they can form pressure maxima and promote both dust accumulation and growth.
Since the sublimation of a volatile is gradual due to opacity changes, a snow line in a protoplanetary disk is in fact a radially extended “snow region” of constant temperature.
{It has been shown that dust can influence disk viscosity through the adsorption of charged particles, and even a minor perturbation in the gas can trigger the excitation of multiple small-scale Rossby vortices.}
Here, we investigate the possibility of Rossby vortex excitation and rapid planetesimal formation at temperature substructures associated with the snow regions, using global 2D gas-dust coupled hydrodynamic simulations that include dust feedback and self-gravity.
We find that an initial temperature substructure in a protoplanetary disk can seed a rapid cascade of long-lived, self-sustaining Rossby vortices.
The vortices accumulate significant amount of dust and the local conditions are favorable for SI as well as gravitational collapse.
However, the vortex formation via this mechanism requires sufficient decoupling between dust and gas, and such conditions may not be met early on when the disk is gas-rich, resulting in a delayed onset of vortex formation. 
The self-sustaining Rossby vortices offer exceptionally favorable conditions for dust growth and the formation of planetesimals, as well as a possible pathway for the rapid formation of planetary cores.
   }

   \maketitle
%

\section{Introduction}

During the process of star formation, a rotating circumstellar disk is inevitably formed from the conservation of angular momentum of the infalling cloud core.
In later stages when the envelope is depleted, the protoplanetary disk can be observed directly, and it is considered to be the cradle for newborn planets \citep{Safronov72}.
With the advent of better observational techniques and instruments, we have over 3200 confirmed exoplanets to date\footnote{https://exoplanetarchive.ipac.caltech.edu/}, and it is likely that most stars host a planetary system \citep{Cassan+12, DC13}.
The observed ubiquity and diversity of exoplanetary systems was not expected, and these observations are challenging current theories of planet formation.
According to the widely accepted core accretion paradigm, planet formation is a bottom-up process, where the submicron-sized dust particles grow into kilometer-sized planetesimals \citep{Pollack+96,Ida-Lin04,Mordasini+08}.
These planetesimals grow further via collisions or pebble accretion and lead to formations of planetary embryos, terrestrial planets, and giant planet cores \citep{Kobayashi+11,Lambrechts-Johansen14}.
Further processes, such as oligarchic growth and runaway gas accretion, take place before the dissipation of the natal gas disk, whereas the final assembly of a planetary system takes several hundred million years \citep{Pollack+96,Kokubo-Ida02,Hansen09,Izidoro+14}.

The submicron-sized interstellar dust starts to grow at the earliest stages of disk formation during cloud collapse in embedded Class 0/I stages \citep{Tsukamoto+23}.
However, the path of dust growth is not without hurdles, and due to the complexity of the physics involved and observational constraints, the processes involved are still not well understood.
Certain barriers prevent dust growth in the environment of protoplanetary disks; this is especially true for dust growth from meter-sized boulders to gravitationally bound planetesimals \citep[for recent reviews, see][]{Drazkowska+23,Birnstiel24}.
One of these obstacles is the drift (or meter-sized) barrier, wherein the particles of this size experience significant aerodynamic drag.
In a typical protoplanetary disk, such particles end up drifting inward and exiting the disk in a relatively short time, leaving the disk devoid of building material for planets \citep{Weidenschilling77}.
Additional barriers to dust growth pertain to the outcome of collisions of dust aggregates, which may depend strongly on their internal composition.
During the collisional evolution of dust, the relative velocity between particles can result in their fragmentation instead of their sticking (fragmentation barrier), while bouncing collisions between aggregates (bouncing barrier) or repulsion between charged grains may also inhibit dust growth (charging barrier) \citep{Okuzumi09, Zsom+10, Birnstiel+12}.

Certain regions of protoplanetary disks offer special conditions for preferential dust growth, which help overcome the aforementioned barriers, and offer a pathway toward planet formation.
High resolution observations of disks in dust continuum emissions by the Atacama Large Millimeter/submillimeter Array (ALMA) show the ubiquitous presence of substructures, such as rings, spirals, and horseshoes \citep{Andrewsetal2018,Dullemond+18,Huang+18}. 
Local pressure maxima may be formed in disk substructures, and, as the dust drifts toward the pressure gradient, such pressure maxima act as particle traps, offering ideal sites for the accumulation and growth of dust.
Once the dust-to-gas ratio is sufficiently high, streaming instability (SI) may be triggered, which is capable of spontaneously clumping and further concentrating the solids into self-gravitating planetesimals \citep{Youdin-Goodman05,Johansen+07}.
The snow or ice line of volatiles and especially water have long been proposed to play an important role in this process \citep{Stevenson-Lunine88,Kretke-Lin07,Drazkowska+17}. 
A snow line can promote planet formation via multiple mechanisms, for example, an increase in solid surface density beyond the snow line, formation of pressure maxima to act as dust traps, "cold finger" effects where the diffusive flow of vapor causes dust enhancements, and effects of ices on grain stickiness \citep{Pontoppidan+14,Oberg-Bergin21}.   
Additionally, the structure of a protoplanetary disk with respect to snow lines may be much more complex.
Using a coupled dynamical and thermodynamical disk model with a temperature-dependent opacity that accounts for the sublimation of dust, \cite{Baillie+15} showed that a snow line is in fact a radially extended "snow region."
The changes in opacity lead to the self-regulation of temperature, whereby the disk dynamically adjusts its heating and cooling balance, resulting in the formation of a snow region rather than a sharp transition at a snow line.

In this study, we investigate the possibility of vortex formation at the water snow region of an MMSN (minimum mass solar nebula) protoplanetary disk.
Small-sized dust grains tend to adsorb electrons and ions, and this process can strongly reduce the conductivity of the gas in their vicinity \citep{Sano+00,Ilgner-Nelson06,Balduin+23}.  
{Magnetorotational instability (MRI) is an important mechanism behind turbulent viscosity in a protoplanetary disk, although additional sources of angular momentum transport, such as magnetic winds and other hydrodynamic instabilities, are simultaneously at work \citep{Lesur+23}.}
The presence of ionized gas is essential for MRI, as it enables the necessary coupling between the gas and the magnetic field. \citep{Balbus-Hawley98}.
The adsorption of charges by dust particles can thus inhibit MRI, meaning that the disk viscosity is reduced in the region of dust accumulation.
This gives rise to a positive feedback cycle involving initial gas perturbation, dust accumulation, reduced local viscosity, and more gas accumulation, thereby amplifying the initial perturbation. 
Linear perturbation analysis or 1D simulations of disk evolution show that such conditions lead to viscous ring-instability (VRI), resulting in the spontaneous formation of concentric rings \citep{DullemondPenzlin2018,Regaly+21}.
However, when non-axisymmetry is allowed, these rings become Rossby unstable upon further evolution and form vortices \citep{Regaly+21,Regaly+23}. 
In this study, we construct an initial equilibrium model of the disk that is consistent with the temperature substructure caused by water snow line.
We evolve this disk adiabatically with the adaptive, dust-dependent viscosity formulation and explore the phenomenon of the formation of vortex cascades and their implications for planetesimal formation.

The structure of this paper is as follows. In Section \ref{sec:hydro}, we explain the hydrodynamical model in brief and describe the initial conditions in Section \ref{sec:init}. 
The main results are presented in Section \ref{sec:results}, where we describe the vortex cascade qualitatively (Section \ref{sec:voertexform}) with the results interpreted quantitatively with respect to their suitability for dust growth and planetesimal formation to follow (Section \ref{sec:voertexevo}).
In Section \ref{sec:constsize} we elaborate on the concept of delayed onset and, lastly, in Section \ref{sec:con} summarize our findings.

\section{Methods}
\label{sec:methods}

\subsection{Hydrodynamic model}
\label{sec:hydro}

We conducted global hydrodynamic simulations of protoplanetary disks using code GFARGO2 \citep{RegalyVorobyov2017a, Regaly20}, which is the GPU-supported version of the code FARGO \citep{Masset2000}, with additional physics modules, e.g., for modeling disk self-gravity, energy equation, and dust species with full backreaction.
The modeling tool set employed in this study is described in \citet{Regaly+21}, and here we provide a summary of its key components.
The 2D evolution in the thin-disk limit is modeled using coupled gas–dust equations, with the dust component modeled as a pressureless fluid, a reasonable approximation for particles with Stokes numbers below unity.
For the dynamical evolution of the gas, we solved the continuity equation, the momentum equation, and the equation for energy transport. An ideal gas equation of state is assumed with vertically integrated pressure, $P = (\gamma-1)\epsilon$, where $\epsilon$ is the internal energy and $\gamma$ is the adiabatic index of the gas.
In the energy equation, viscous heating is considered explicitly, while remaining disk heating and cooling processes are considered implicitly via a $\beta$-cooling prescription, which maintains the initial temperature structure.
In addition to the stellar gravitational potential, contribution from the disk self-gravity as well as gravitational effects from indirect potential, arising from the movement of the star around the common barycenter of the system, are considered.
For all gravitational calculations, contributions from both gas and solid components is taken into account. 
The disk's self-gravity is calculated without gravitational softening, by solving the Poisson integral by using the 2D Fourier convolution theorem for polar coordinates logarithmically spaced in the radial direction \citep{RegalyVorobyov2017a}.
{It is possible that 2D methods can overestimate gravitational forces without the correct value of the smoothing length; however, our method is independent of this parameter and produces comparable results \citep{Muller+12,Vorobyov+24}.}
For the dust component, the continuity and momentum equations are solved, while the turbulent diffusion of solid is modeled by the gradient diffusion approximation.
The back-reaction term due to dust drag is taken into account, and the simulations are performed under the assumption of either a constant Stokes number or a constant dust grain size.

We use the $\alpha$-prescription of \cite{ShakuraSunyaev1973} to model turbulent viscosity from MRI, modified such that the local viscosity depends on the dust and gas surface densities, 
\begin{equation}
    \nu=\alpha_\mathrm{bg}c_\mathrm{s}h\left(\frac{\Sigma_\mathrm{d}}{\Sigma_\mathrm{d, 0}}\right)^{\phi_\mathrm{d}}\left(\frac{\Sigma_\mathrm{g}}{\Sigma_\mathrm{g,0}}\right)^{\phi_\mathrm{g}},
    \label{eq:visc}
\end{equation}
where $c_s$ is the sound speed, and $h$ is the gas scale height.
Here $\alpha_\mathrm{bg}$ is the maximum possible background viscosity, while $\Sigma_\mathrm{g,0}$ and $\Sigma_\mathrm{d,0}$ are the initial gas and dust densities \citep{DullemondPenzlin2018, Regaly+21}.
In this parametric viscosity prescription, the exponents $(\phi_\mathrm{d},\phi_\mathrm{g})$ specify how the dust concentration alters the local viscosity in the disk.
With the assumption that the dust particles adsorb charges and reduce the gas-magnetic field coupling \citep{Sano+00,Ilgner-Nelson06,Balduin+23}, $\alpha$ is inversely proportional to dust-to-gas ratio, with $(\phi_\mathrm{d},\phi_\mathrm{g}) = (-1, 1)$.
Nonideal MHD simulations suggest the presence of ``zonal flows" in the disk, wherein the vertical magnetic flux is strongly enhanced (or diminished) in the low (or high) gas density regions \citep{Johansen09,Bai15}.
As a result, the effective viscosity increases in the low density gaps and is suppressed in the denser regions.
We can mimic the associated variations in viscosity with $(\phi_\mathrm{d},\phi_\mathrm{g}) = (-1, -1)$. 
For more details on the equations and code setup, see \cite{Regaly+21}.

\subsection{Initial conditions}
\label{sec:init}

In this study, we focused on the water snow region, which forms a plateau at its condensation temperature of 160 K.
According to the model of \cite{Baillie+15}, which considers dynamical and thermodynamical disk evolution with a realistic temperature-dependent opacities, this plateau is located between 1.5 and 2.5 au for a sun-like star, after 1 Myr evolution of an MMSN disk.
The mechanism behind formation of a constant temperature plateau is related to gradual sublimation water due to opacity changes and can occur for any volatile species including refractory material in the dust grains \citep{Woitke+24}.
These findings form the basis for initial conditions for our hydrodynamic simulations.
{ Although planet formation possibly begins before the Class II stage, when the disk is more massive \citep{Tychoniec+20}, our initial conditions follow the well-established standards of MMSN, which offer a simple and representative starting point.} 
Note that for simplicity, the disk is built such that only the water snow region is represented, although additional temperature substructures may exist in a realistic protoplanetary disk (see Fig. 10 of \citealp{Baillie+15})

\begin{figure}
    \centering
    \includegraphics[width=0.98\columnwidth]{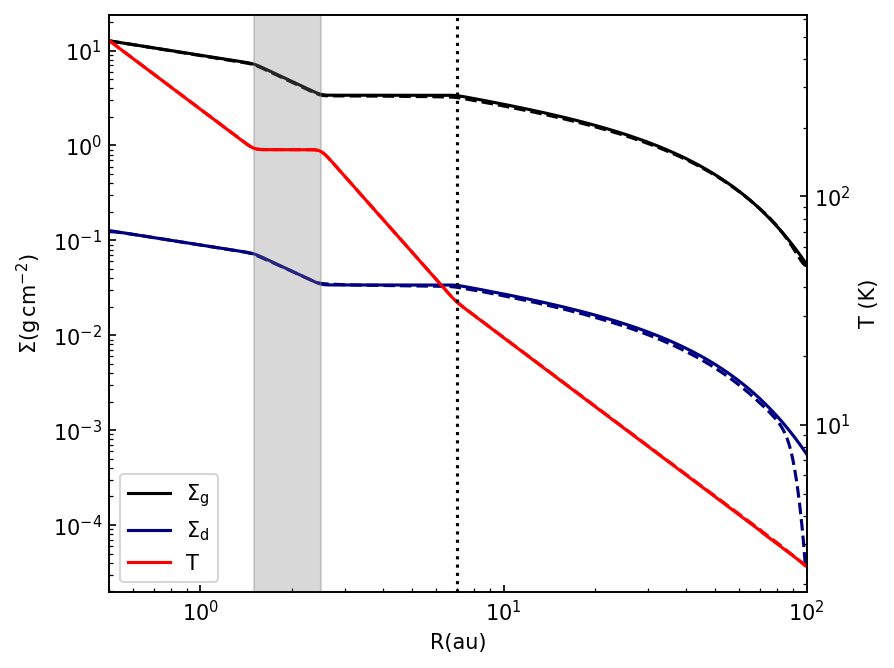}
    \caption{Azimuthally averaged profiles of the disk gas and dust surface density as well as the midplane temperature. The disk is evolved adiabatically with a constant $\alpha=10^{-3}$ to show the pseudo-equilibrium of the initial configuration (solid lines) after 50 kyr of evolution (dashed lines). The gray band shows the water snow region. The dotted line marks $R_{\rm end}$, outside of which an exponential taper is applied.}
    \label{fig:init}
\end{figure}

To construct the initial disk structure, we used the standard viscous accretion disk theory, in which the gas surface density and temperature profiles are given by $\Sigma_{\rm g}(R) \propto R^p$ and $T(R) \propto R^q$ \citep{APiA}.
The assumption of constant accretion rate yields, $p+q = -3/2$.
With this simple relation, we obtain the initial steady-state configuration of the disk as follows.
For the disk inside of the plateau at $r<1.5$ au, $p=-1/2$ and $q=-1$. 
At the water snow region between $1.5$ and $2.5$ au, $p=-3/2$ and $q=0$, so as to obtain a constant temperature plateau. 
Beyond 2.5 au the temperature decreases more steeply, with, $p=0$ and $q=-3/2$, until it intersects the original $q=-1$ line at $R_{\rm end}$. 
This point onward, the original $\Sigma_{\rm g}(R)$ profile continues, but with an exponential taper to the gas surface density, with a characteristic radius of $R_{\rm c} = 50$ au, i.e.,
\begin{equation}
    \Sigma_\mathrm{g}(R)=\Sigma_\mathrm{g}(R_\mathrm{end}) \left(\frac{R}{R_\mathrm{c}}\right)^{p}\exp{\left(-\frac{R}{R_\mathrm{c}}\right)^{2-p}}.
\end{equation}
An exponential decay at the outer boundary is observed in such gas tracers as rotational lines of CO \citep{WC11}, and it is also necessary in the simulations to avoid gravitational instability in the outer disk from self-gravity.
The $\Sigma_\mathrm{g}$ and temperature profiles for the entire disk are thus piecewise continuous functions of the radius.
Additionally, a Gaussian convolution with a full width at half maximum of 5 cells is used to smooth the intersections of these segments, where the temperature derivative changes, so as to avoid kinks and associated artifacts during the disk evolution.
The gas surface density at 1 au, $\Sigma_\mathrm{g}(1 {\rm au})$, is set such that the disk with this aforementioned $\Sigma_\mathrm{g}$ profile has a mass of 0.01 $M_\odot$.
The dust surface density is obtained from the gas profile by multiplying it with the assumed initial dust-to-gas ratio ($\zeta_{d2g, 0}$).
The resultant gas, dust and temperature profiles are plotted in Fig. \ref{fig:init}.
To test the stability of the disk, we evolved this configuration adiabatically for 100 orbits at 1 au, for a disk with a constant $\alpha=10^{-3}$ and $\zeta_{d2g, 0}=0.01$. 
The final profiles are also shown in Fig. \ref{fig:init}, which imply that such a disk with a temperature substructure is indeed in pseudo-equilibrium.

The values of physical parameters for the fiducial disk as well as some important quantities are listed in Table \ref{tab:fiddisk}. 
These values remain unchanged across various simulations, with the exception of a low mass disk, as explained later in this section. 
The cooling parameter $\beta$ is chosen to be 0.1; however, a larger value of 1.0 does not change the results of our simulations qualitatively.
The resolution choice attempts to strike a balance between computational cost and resolving the phenomenon at hand. With a logarithmic grid in the radial direction, the radial resolution at a given point is approximately one tenth of the local scale-height.  
The results presented in this study remain unaltered at a higher resolution, although we do not present a convergence test in the interest of conciseness.

\begin{table}
    \caption{Fiducial disk parameters.}
    \centering
    \begin{tabular}{l|r}
\hline
         Physical parameter & Value \\
\hline
         Disk mass~$(M_\odot)$, $M_\mathrm{disk}$ & 0.01 \\
         Stellar mass~$(M_\odot)$, $M_\mathrm{star}$ & 1.0\\
         Cutoff radius~(au), $R_\mathrm{c}$ & 50 \\
         Surface density at 1 au $(\mathrm{g\,cm^{-2}})$, $\Sigma_\mathrm{g}(1 {\rm au}) $ & $8.88$\\
         Gas scaleheight, $h$ & $0.05$\\
         Adiabatic index of gas, $\gamma$ & $1.4$\\
         Cooling parameter, $\beta$ & 0.1\\
         Inner disk radius~(au), $R_\mathrm{in}$ & 0.5 \\
         Outer disk radius~(au), $R_\mathrm{out}$ & 100 \\
         Radial resolution, $N_\mathrm{R}$ & 768 \\
         Azimuthal resolution, $N_\mathrm{\Phi}$ & 896 \\
\hline
    \end{tabular}
    \label{tab:fiddisk}
\end{table}

\begin{table*}
\caption{\bf List of simulations.}
\label{table:sims}
\begin{tabular}{|l|}
\hline
\begin{tabular}{p{2.4cm}p{0.8cm}p{1.1cm}p{1.1cm}p{0.8cm}p{2cm}p{2.5cm}p{2cm}p{2cm}}
\hspace{-0.1cm} Model Name &  $ \phi_{\rm d}, \phi_{\rm g}$  &  St / Size  & $\zeta_{\rm d2g,0}$ &  $\alpha_{\rm bg}$  & $M_{\rm disk} (M_\odot)$ & \hspace{-0.5cm}{Significance} &  Outcome \\
\end{tabular}\\ \hline
\begin{tabular}{l}
$\kern-\nulldelimiterspace\left.
\begin{tabular}{p{2.4cm}p{0.8cm}p{1.1cm}p{1.1cm}p{0.8cm}p{1.2cm}p{2.8cm}p{2cm}p{2cm}}
1. \simname{Pl\_STD}   &  -1,1  & 0.01 &  0.01 &  $10^{-3}$  &  0.01 &   Dust adsorption & Vortices \\
2. \simname{Pl\_DG}   &  -1,1  & {0.1} &  0.01 &  $10^{-3}$ & 0.01 & Dust growth & Vortices \\
3. \simname{Pl\_SM}   &  -1,1  & {0.001} &  0.01 &  $10^{-3}$ & 0.01 & Coupled small dust & No vortices \\
4. \simname{Pl\_ALPHA}   &  -1,1  & 0.01 &  0.01 &  ${10^{-2}}$  & 0.01 & Higher disk $\alpha$ & No vortices \\
5. \simname{Pl\_MHD}   &  {-1,-1}  & 0.01 &  0.01 &  $10^{-3}$ & 0.01 & Gas MHD effects & Vortices+bands \\
6. \simname{Pl\_Z}   &  -1,1  & 0.01 &  {0.05} &  $10^{-3}$  & 0.01 &  Higher metallicity & Vortices \\
\end{tabular}\right\}$ Constant Stokes  
\\ 
\end{tabular}  
\\ \hline
\begin{tabular}{l}
$\kern-\nulldelimiterspace\left.
\begin{tabular}{p{2.4cm}p{0.8cm}p{1.1cm}p{1.1cm}p{0.8cm}p{1.2cm}p{2.8cm}p{2cm}p{2cm}}
7. \simname{Pl\_1mm}   &  -1,1  & {1 mm} &  0.01 &  $10^{-3}$  & 0.01 & Constant dust size & Vortices   \\
8. \simname{Pl\_0.1mm}   &  -1,1  & {0.1 mm} &  0.01 &  $10^{-3}$ & 0.01 & Small-sized dust & No vortices \\
9. \simname{Pl\_0.1mm\_LM}   &  -1,1  & 0.1 mm &  0.01 &  ${10^{-3}}$ & {0.001} & Lower mass disk & Vortices  \\
  \end{tabular}\right\}$ Constant size
\end{tabular}  
\\
\hline
\end{tabular}\\
\end{table*}

For this study, we considered a total of nine simulations, as listed in Table \ref{table:sims}.
The first set of simulations is conducted with the assumption of constant Stokes number of the dust species, while the second set assumes a constant particle size. 
An assumption of internal density of dust particles, e.g., $\rho_s=1.6~{\rm g\,cm^{-3}}$, in addition to the gas surface density, makes it possible to calculate the dust size for constant Stokes number models, and vice versa.
The first simulation in the list is a standard or fiducial model, \simname{Pl\_STD}, which will be used to demonstrate the vortex formation at the temperature plateau of water snow line in detail.  
{ Recent observational as well as theoretical evidence suggests that the turbulence within typical protoplanetary disks is about an order of magnitude lower than the typical assumption of $\alpha \approx 0.01$ \citep{Kadam+25,Villenave+25}.
Hence, we chose $\alpha_{\rm bg}=10^{-3}$ as the standard value for a fully MRI-active disk driven by turbulent viscosity.}
In the subsequent simulations, we used model \simname{Pl\_STD} as a benchmark and change one model parameter at a time, to investigate the effects of different disk conditions as well as dust particle properties.
In each row, the parameters that are changed are highlighted in bold for clarity. Their significance and the simulation outcome are also listed. 
This limited parameter search provides us with a comprehensive understanding of the vortex cascade formation at water snow region.

\section{Results}
\label{sec:results}

\subsection{Vortex formation at the water snow region}
\label{sec:voertexform}

\begin{figure*}
\begin{tabular}{l}
\hspace{2.6 cm} 1.9 kyr \hspace{2.5 cm} 2.0 kyr \hspace{2.7 cm}  6.9 kyr  \hspace{2.5 cm}   13.0 kyr\\
\vspace{-0.3cm}\includegraphics[width=\textwidth]{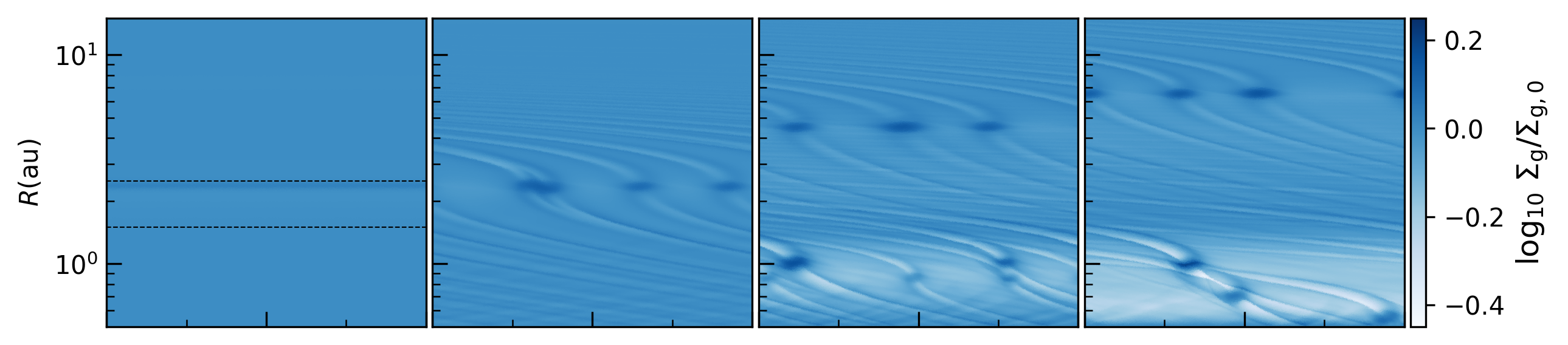} \\
\vspace{-0.3cm}\includegraphics[width=\textwidth]{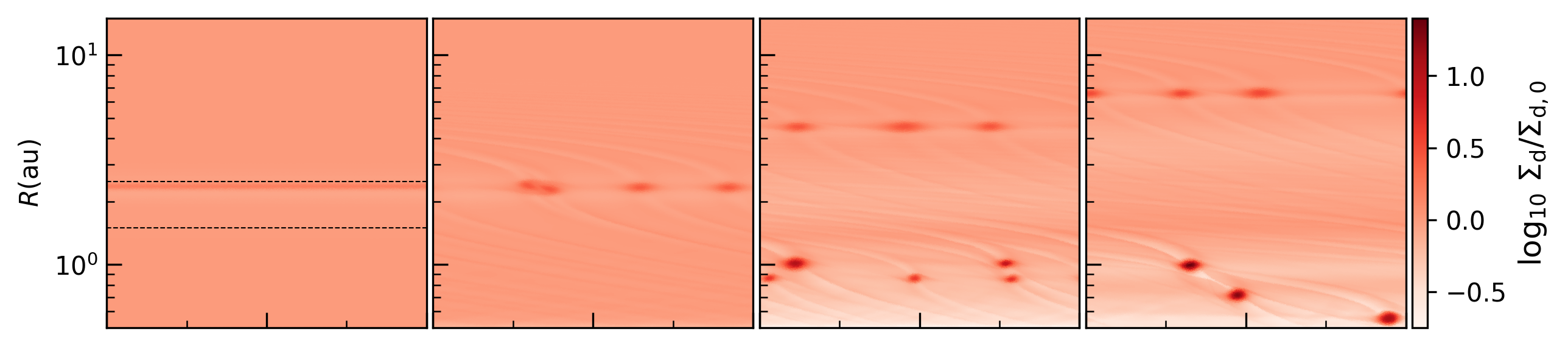} \\  
\includegraphics[width=\textwidth]{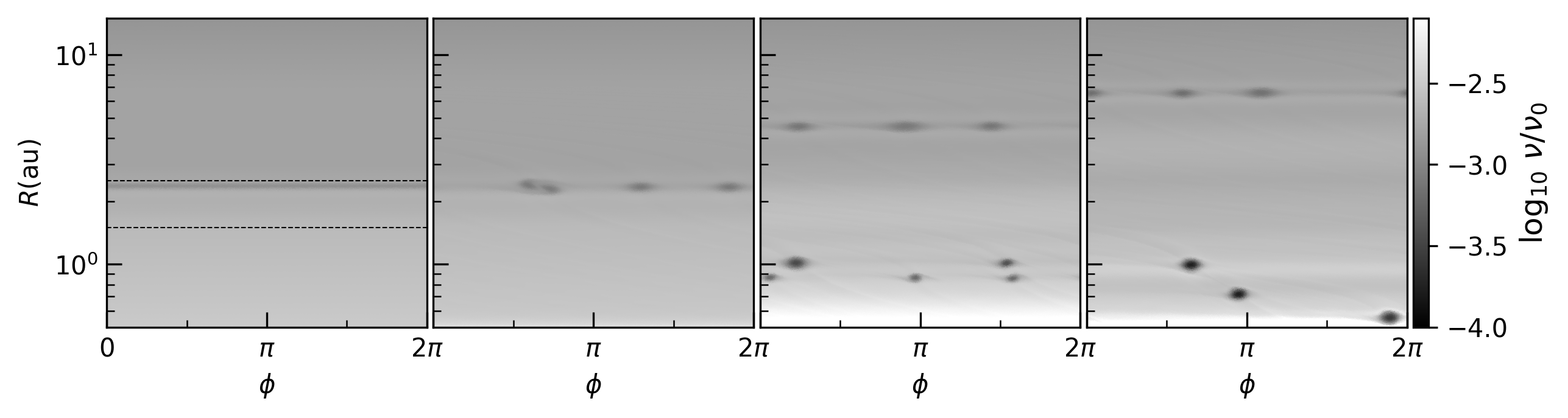} \\
\end{tabular}
\caption{Progression of a self-sustaining vortex cascade in an MMSN disk for model \simname{Pl\_STD}, showing the distribution of normalized dust, gas, and viscosity. The dashed lines show the location of the initial temperature plateau corresponding to the water snow region.}
\label{fig:evo2d}
\end{figure*}

\begin{figure*}[h]
\centering
\includegraphics[width=0.75\textwidth]{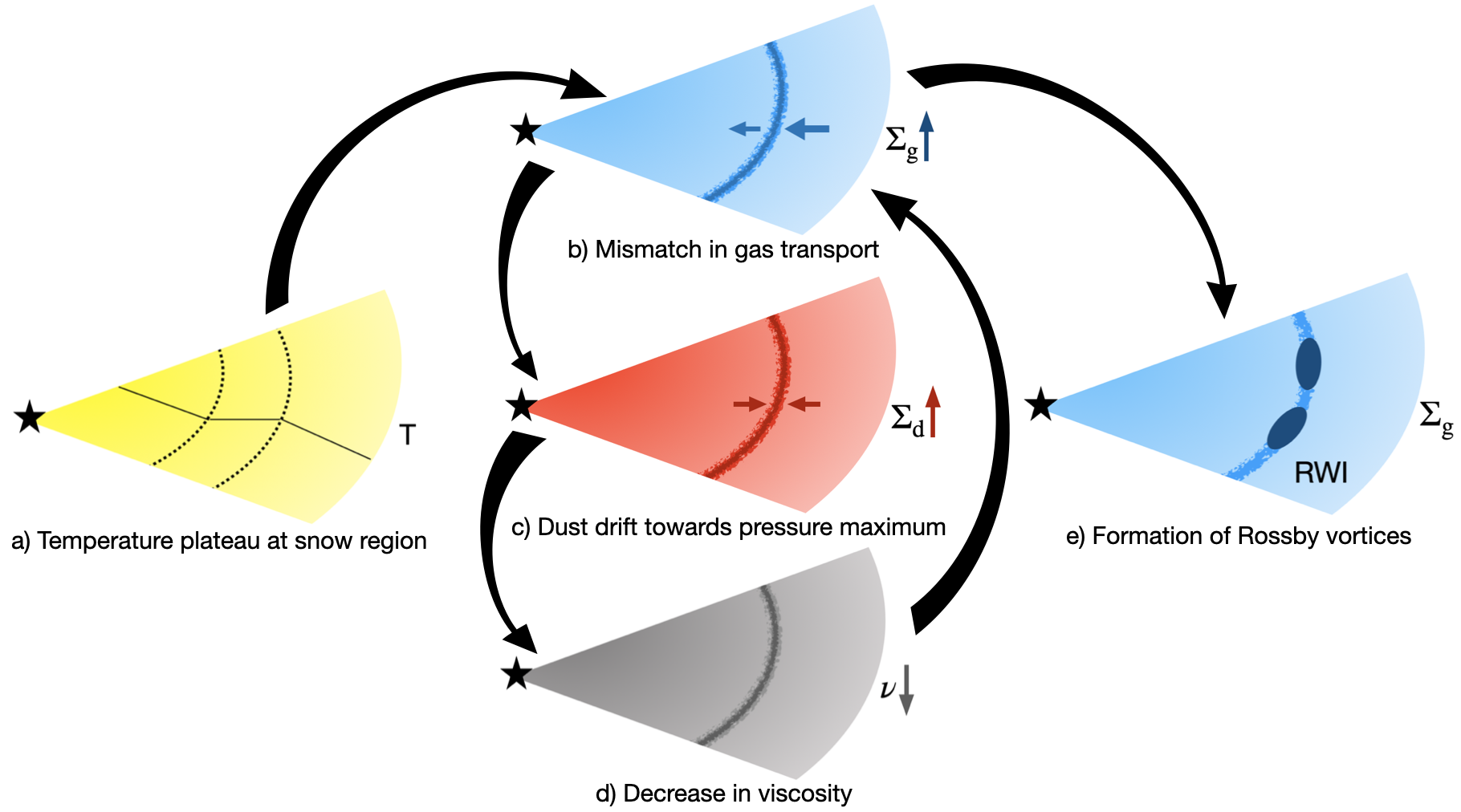} \\
\caption{Cartoon explaining the mechanism behind the formation of self-sustaining vortices at the temperature plateau at the water snow region. The temperature substructure gives rise to marginal mismatch in the gas accretion rates, seeding the initial perturbation in gas surface density. The dust is attracted toward the resulting pressure maximum, which, in turn, lowers the local dust-dependent viscosity. This, in turn, creates a bottleneck to angular momentum transport, amplifying the perturbation in the gas and thus leading to a positive feedback loop.
In a 1D disk, such feedback leads to VRI, forming multiple concentric rings.
In 2D simulations, the gradient of gas surface density eventually becomes sharp enough to trigger Rossby wave instability, and vortices are formed. The positive feedback responsible for VRI contributes toward the stability of formed vortices.
}
\label{fig:schematic}
\end{figure*}

In this section, we demonstrate how the vortices form at the water snow region, their evolution, typical behavior, and qualitative properties. 
Previously, we showed that with the assumption of dust-dependent viscosity, a cascade of small-scale, self-sustaining vortices ensues with an initial Gaussian perturbation in gas surface density \citep{Regaly+21}.
However, such ad hoc initial conditions are rather artificial, and here we show that similar vortex cascade is brought forth by the water snow region, which naturally forms in a disk with solar composition. 
Fig. \ref{fig:evo2d} shows the evolution of gas and dust surface densities, as well as viscosity for simulation \simname{Pl\_STD}.
The quantities are normalized with respect to the initial value to enhance the substructure and because they are used in the calculation of the disk viscosity (Eq. \ref{eq:visc}).
The snow region between 1.5 and 2.5 au is marked with dashed lines in the first panel of each row. 
The outer edge of the snow region first forms a ring in gas, which accumulates dust in its pressure maximum, and this in turn results in a decrease in the local viscosity.
The positive feedback cycle that follows is similar to viscous ring instability (VRI), wherein dust accumulation lowers viscosity, which in turn leads to gas accumulation; the resulting enhancement in pressure maximum attracts even more dust \citep{DullemondPenzlin2018}. 
Since the viscous timescale is smaller closer to the star, one might expect the initial ring to form at the inner edge of the temperature plateau.
However, the pressure maximum at the outer edge has a much larger reservoir of dust entering from the outer disk, which explains why the initial ring forms consistently at 2.5 au.
Both linear perturbation analysis as well as 1D simulations suggest that VRI gives rise to several concentric rings in the disk \citep{DullemondPenzlin2018,Regaly+21}.
Simulations in 2D show that the steep gradients developed during this process make the ring Rossby unstable upon further evolution.
These vortices are termed self-sustaining because the feedback between increased dust concentration and reduced viscosity, which gives rise to VRI, also ensures their secular stability.
Fig. \ref{fig:schematic} shows a schematic of this process.
Note that several other mechanisms resulting from unusually high concentrations of dust, including SI, are capable of destabilizing or dismantling vortices \citep{Fu+14,Raettig+15,Lovascio+22}.

As seen in the second column of Fig. \ref{fig:evo2d}, a ring typically breaks into several small-scale vortices.
{ These vortices are different from the "large-scale" vortices that occupy a significant extent of the disk in the azimuthal direction and may be responsible for horseshoe-shaped brightness asymmetries seen in the millimeter-wavelength \citep{Regalyetal2012,RegalyVorobyov2017a,Regaly+23}.}
While evolving, the vortices show complex behavior such as increasing strength in terms of dust accumulation, inward migration and merging.
{ The Rossby vortices interact viscously and gravitationally with the disk, driving angular momentum transport.
Similar to a low-mass planet embedded within a gaseous disk, this interaction produces large-scale spiral density waves originating at the vortices, which are clearly seen in gas distribution.
The resulting torques cause their orbital decay and the vortices typically migrate inward from their radial location of birth.}
{  A significant consequence of this process is that, when multiple vortices are present, numerous large-scale spiral waves may interfere constructively at different radii, either inside or outside their radial location. 
The resulting perturbations in gas surface density produce pressure maxima that can attract dust and give rise to similar positive feedback between dust and viscosity, thus forming a new generation of vortices. }
Such cascades are shown in the subsequent panels of Fig. \ref{fig:evo2d}, which can extend at much larger radii than the initial location of their origin or, in this case, the snow region.

\begin{figure*}
\centering
\includegraphics[width=0.68\textwidth]{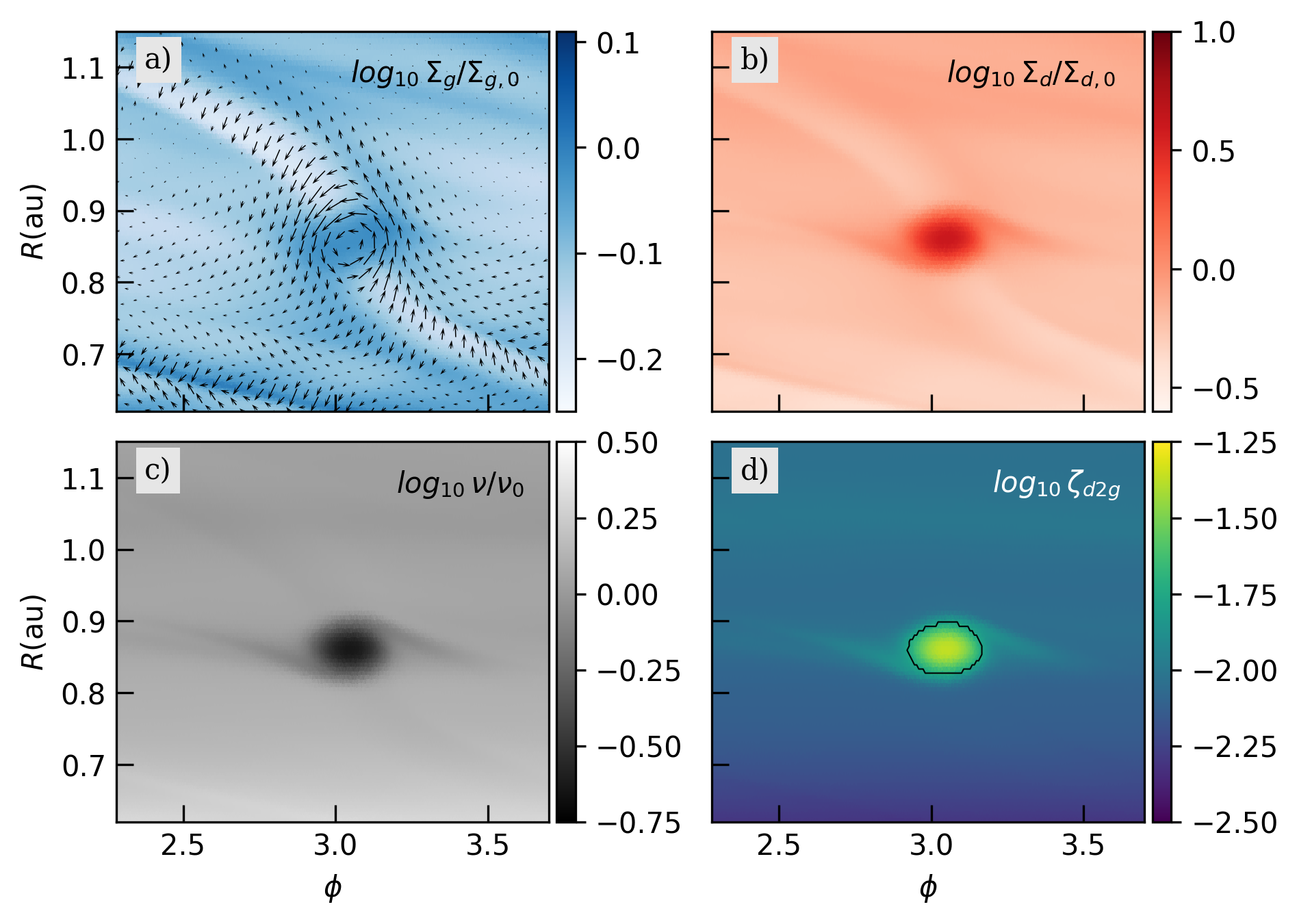} \\ 
\caption{Enlarged view of an isolated vortex in model \simname{Pl\_STD}, showing the normalized gas and dust surface densities, as well as the viscosity and dust-to-gas surface density ratio.
The arrows in the first panel show the gas velocity field.
The contour in the last panel encompasses the region prone to SI. 
}
\label{fig:zoom2d}
\end{figure*}

Fig. \ref{fig:zoom2d} focuses on an individual isolated vortex formed at about 0.8 au in \simname{Pl\_STD} simulation. 
Panel a) shows the normalized gas density, along with the gas velocity field; the latter is the difference between the gas velocity and the local Keplerian velocity.
The disk rotates in a clockwise direction and the anticyclonic nature of the vortex is clearly seen in the velocity field.
For the Stokes number of 0.01 used in this simulation, the dust is relatively well coupled to the gas, and its velocity appears similarly curled.
The strong accumulation of the dust is seen in panel b) at the eye of the vortex and the viscosity shown in panel c) mirrors the dust concentration. 
The last panel d) shows the dust-to-gas surface density ratio, which shows significant enhancement over the background value of 0.01.
If the dust accumulation is sufficiently strong, SI may cause the dust to spontaneously clump to form gravitationally bound planetesimals.
The canonical criterion for the onset of SI is considered to be enhancement of the midplane dust to gas volumetric density over unity \citep{Youdin-Johansen07}.
However, we considered the empirical relation derived by \cite{Yang+17} to identify the regions of the disk prone to SI: 
\begin{equation}
    {\rm log}(\zeta_{\rm d2g}) = 0.1({\rm log}\, {\rm St})^2 + 0.2 {\rm log}\,{\rm St} -1.76,
    \label{eq:SI}
\end{equation}
which is a more robust predictor.
{ Note that although vortices offer an efficient mechanism for concentrating dust and its growth, the exact conditions for dust clumping and SI onset in vortical flows are not known.}
The contour in the last panel encompasses the region susceptible to SI, { calculated using the criterion described in Eq. \ref{eq:SI}.} 
The core of a typical vortex accumulates sufficient dust to trigger SI and, thus, provides an ideal site for planetesimal formation.
The cascade of vortices implies that planetesimal formation can occur over a large radial extent of the disk. 
Three-dimensional simulations of SI suggest that it is difficult to form planetesimals in axisymmetric pressure traps, unless the dust is already grown to centimeter-sized pebbles \citep{Carrera-Simon21,Carrera-Simon22}. 
The localized nature of the vortex core offer a possible alternative for planetesimal formation via SI, as well as for their further growth to form seeds of planetary embryos.

\subsection{Vortex evolution and planetesimal formation}
\label{sec:voertexevo}

\subsubsection{Effects of dust-gas coupling}

\begin{figure}
\centering
\includegraphics[width=0.45\textwidth]{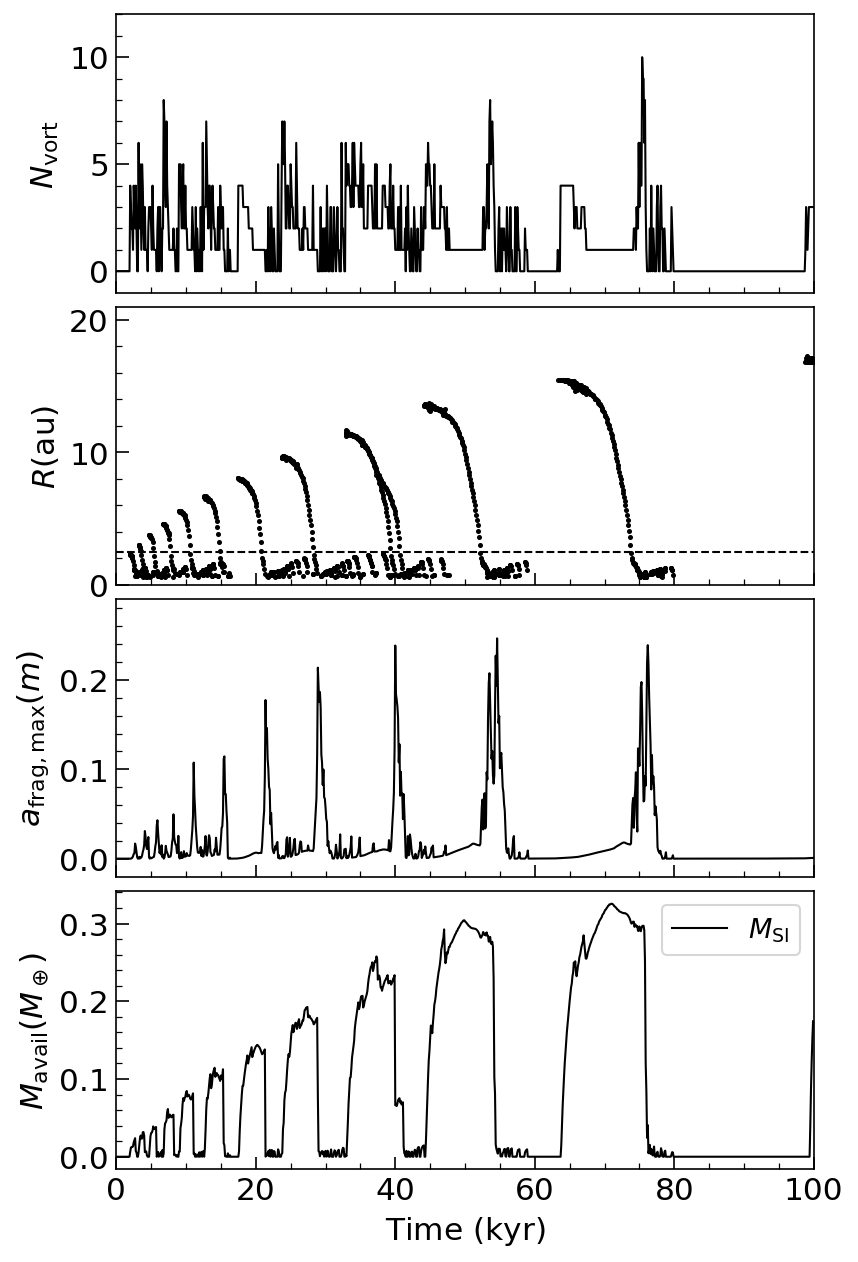} \\ 
\caption{Temporal evolution of the aggregate properties of the vortices formed in model \simname{Pl\_STD}, showing the number of vortices, their position, the maximum fragmentation size of the dust grains, and the mass available for SI.}
\label{fig:Pl_STD}
\end{figure}

In this section, we investigate the temporal evolution of the disk with respect to the vortex cascade and prospects of planetesimal formation.
The focus here is on the effects of dust-gas coupling, which is probed using the first three models in Table \ref{table:sims}. To reiterate, the models \simname{Pl\_STD}, \simname{Pl\_DG}, and \simname{Pl\_SM} have a constant Stokes number of 0.01, 0.1, and 0.001, respectively, and are otherwise identical.
Consider Fig. \ref{fig:Pl_STD} for the standard disk model, where the first panel shows the number of vortices in the disk ($N_{\rm vort}$), and the second panel shows their radial position.
The location of individual vortices and their number is calculated by detecting the local maxima in the field of the dust to gas ratio, similar to the method used in \citet{Regaly+21}.
After the simulation begins, the initial ring forms relatively quickly, within 2000 yr, and it soon breaks up into a number of small-scale vortices. 
An abrupt increase in the number of vortices corresponds to successive generations of vortices, which originate at different radii in the disk.
This is clearly seen in the second panel of the figure in their radial position. 
Initially the vortices start forming at the location of the temperature plateau, marked by the dashed line at 2.5 au.
The successive generations are triggered at progressively larger radii in the disk up to 14 au, giving rise to the snake-like patterns. 
We hypothesize that this behavior is due to two factors. First, the interference of large-scale spiral waves creates local pressure maxima and thus suitable sites for dust accumulation and vortex formation.
Second, the vortices in the inner disk deplete their immediate surroundings of dust, thereby creating a radial gradient in the dust to gas ratio, which is sufficient to trigger vortex formation at slightly larger radial distance. This explains how the vortices are formed at an increasing radius, even in the absence of any other vortices in the disk at that time.
The frequency of vortex occurrence generally decreases with time, presumably because of the gradual depletion of the finite dust reservoir in the disk.
We plan to conduct a separate follow up study focusing on detailed behavior of vortices, in order to answer questions related to triggering mechanisms of successive generations and migration rates of vortices in different disk conditions.

\begin{figure}
\centering
\includegraphics[width=0.445\textwidth]{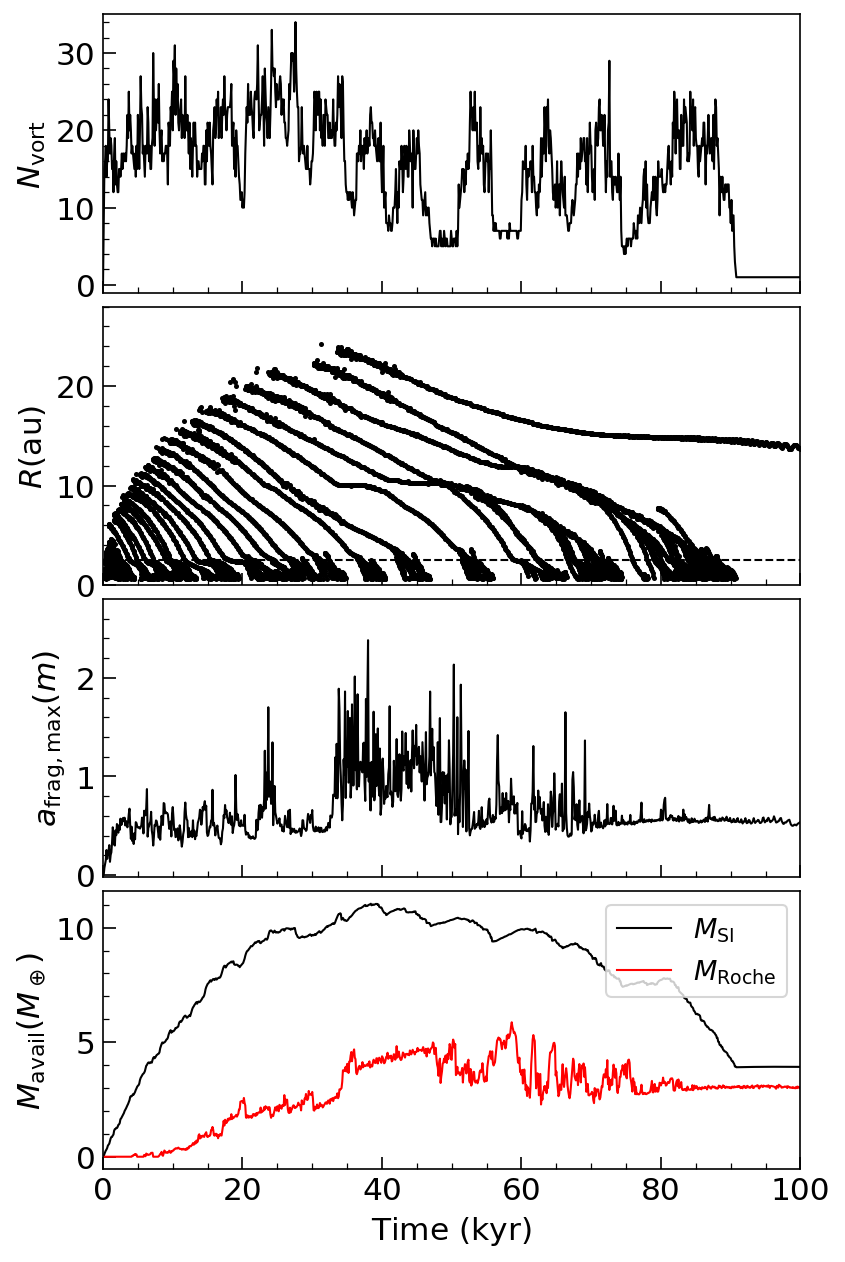} \\ 
\caption{Same as Fig. \ref{fig:Pl_STD}, but for model \simname{Pl\_DG}. The third panel additionally shows mass exceeding Roche density.}
\label{fig:Pl_DG}
\end{figure}

The third panel of Fig. \ref{fig:Pl_STD} shows the maximum turbulent fragmentation size of the dust grains within the disk, which is located at the center of a vortex.
We neglect the contribution from the innermost 0.1 au, in order to avoid any artifacts that may be caused by interactions with the inner boundary.
This fragmentation size is calculated as 
\begin{equation}
 a_{\rm frag}=\frac{2\Sigma_{\rm g}v_{\rm frag}^2}{3\pi\rho_{\rm s}\alpha c_{\rm s}^2},
 \label{eq:afrag}
\end{equation}
where $v_{\rm frag}= 3~{\rm m\,s^{-1}}$ is the fragmentation velocity, $\rho_s=1.6~{\rm g\,cm^{-3}}$ is the internal density of dust particles, and $c_{\rm s}$ is the local sound speed \citep{Birnstiel+12}.
Due to collisions, fragmentation tends to put an upper limit on the dust size, especially in the inner disk.
Since the dust within a vortex drifts at the migration speed of the vortex itself, the turbulent fragmentation limit is more suitable as compared to drift-induced fragmentation barrier.
This size provides an estimate on how large the dust particles can grow as a result of collisional evolution, before the onset of SI.
Due to increased local gas density and lower viscosity, dust pebbles can achieve an approximate maximum size of 20 cm inside of the vortices, which is significantly larger than that typical disk environment.
For a given generation of vortices, more dust accumulates with time, decreasing the local $\alpha$ as well as $c_{\rm s}$ via diminished viscous heating, which results in temporal peaks in $a_{\rm frag}$. 
The last panel shows the mass available in the disk for SI according to the criterion specified in Eq.~\ref{eq:SI}.
Coinciding with different generations of vortices, the mass available for SI increases, until the vortices exit the disk through the inner boundary.  
The vortices are efficient in producing the conditions suitable for triggering SI and their secular stability ensures formation of planetesimals.
High resolution 3D shearing-box simulations of protoplanetary disks suggest that the growth timescale for SI is of the order of 100 times the local orbital period \citep{Simon+16}.
If we assume that SI is induced in vortical flows under similar conditions as in a Keplerian disk, the temporal condition is also satisfied within the vortices in most instances, despite of their inward migration and transient nature.

In the case of model \simname{Pl\_SM}, which has a dust component with a Stokes number of 0.001, no vortices develop in the disk. This indicates that a certain amount of dust drift or dust-gas decoupling is necessary to form vortices with the dust-dependent prescription of viscosity.
Fig. \ref{fig:Pl_DG} shows the evolution of vortices for model \simname{Pl\_DG}, which has dust with a Stokes number of 0.1.
As seen in the first panel, approximately 20 vortices spawn at the location of the snow plateau soon after the simulation begins.
The number of vortices at any given time is much larger than model \simname{Pl\_STD}, which implies that a larger Stokes number is more favorable for vortex formation via the mechanism of dust-dependent viscosity.
The successive generation of vortices develop much more rapidly as well, with several generations being present simultaneously in the disk.
This cascade of vortex formation extends up to about 25 au, which is 10 times farther than the snow region, seeding the entire inner disk with vortices. 
Additionally, the snow region itself continually gives rise to vortices, which exit the disk relatively quickly.  

The third panel of Fig. \ref{fig:Pl_DG} shows the maximum fragmentation size of the dust particles, which permits the formation of meter-sized boulders.
This limit is much higher than that allowed in a typical MMSN protoplanetary disk.
The last panel shows the dust mass available for SI in the disk, which starts increasing as soon as the simulation begins and exceeds 10 Earth masses near 50 kyr. 
Additionally, this panel shows the amount of dust mass that exceeds Roche density at the given radius.
For this, we compared the midplane volumetric density of the dust, calculated using typical assumptions of vertical equilibrium, with the critical Roche density \citep{Chandrasekhar63},
\begin{equation}
    \rho_{\rm Roche} = \frac{9}{4 \pi} \frac{M_{\rm star}}{R^3}.
\end{equation} 
The Roche density is the minimum density required for an orbiting body to resist the gravitational torques of the central star and remain intact, held together by its own gravity.
As seen in Fig. \ref{fig:Pl_DG}, almost half of the mass prone to SI inside the vortices also exceeds the critical Roche density. 
These Roche-stable vortices are typically located farther out in the disk, as the Keplerian velocity decreases with the radial distance.
This implies that individual vortices are able to grow dust up to meter-sized boulders, which subsequently undergo SI to form planetesimals, which may in turn undergo further gravitational collapse to form more massive objects similar to planetary embryos. 
These conditions are exceptionally suitable for rapid formation of planetary systems.
In our limited parameter search for this study, the critical Roche density is exceeded in model \simname{Pl\_DG} only, however, conditions for SI are satisfied in every case where the vortices appear.

\subsubsection{Effects of disk viscosity and emulating nonideal MHD of gas}

Model \simname{Pl\_ALPHA} probes vortex formation in a higher viscosity disk, with $\alpha_{\rm bg} = 0.01$.
It is possible that higher viscosity suppresses vortex formation and negatively impacts its long-term survival, as it tends to smear out any sharp density transitions and and thus the pressure gradients necessary for triggering or maintaining Rossby wave instability \citep{Lovelace-Romanova14}. 
In the case of model \simname{Pl\_ALPHA}, the vortices did not develop, which suggests that a low effective $\alpha$ is necessary for their formation via the mechanism of dust-dependent viscosity. 
Recent observations as well as nonideal magnetohydrodynamic simulations of protoplanetary disks suggest that midplane turbulent $\alpha$ is much lower than canonical value of 0.01 \citep{Pinte+16,Dullemond+18,Lesur21b}.
It is likely that disk winds make a significant contribution in driving disk accretion, which would transport both mass and angular momentum vertically, instead of their redistribution within the disk \citep{Kadam+25}.
Although the effects of varying disk viscosity on the formation of Rossby vortices are relatively well known, the effects of disk winds are complicated and have not been explored in the literature.

\begin{figure}
\centering
\includegraphics[width=0.43\textwidth]{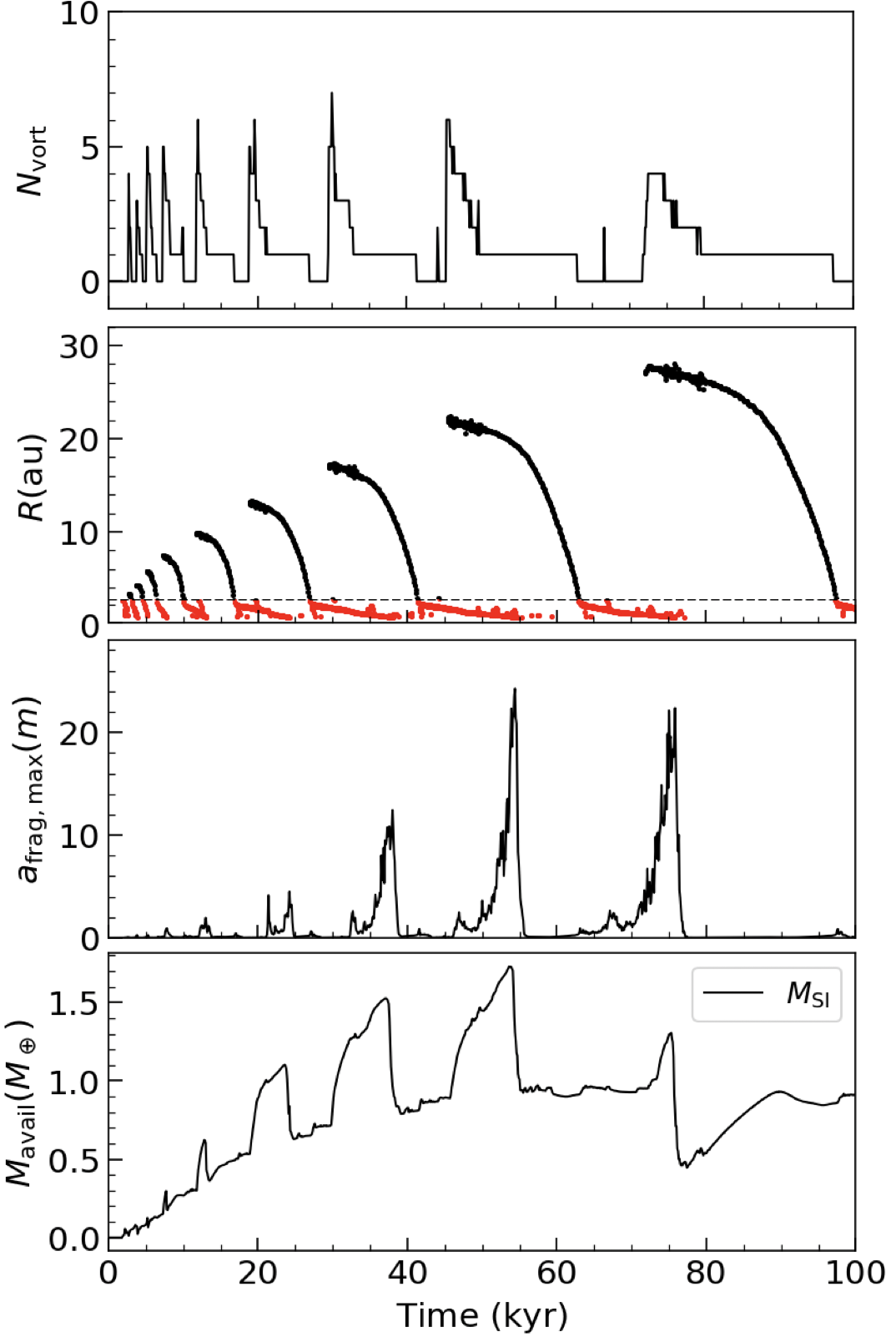} \\ 
\caption{Same as Fig. \ref{fig:Pl_STD}, but for model \simname{Pl\_MHD}. The red color in the second panel indicates the position of dust-gas bands.}
\label{fig:Pl_MHD}
\end{figure}

Local shearing box simulations of protoplanetary disks, which include nonideal MHD effects, suggest formation of zonal flows \citep{Johansen09,Bai15}. These are characterized by an inverse relation between gas density and magnetic field, such that the magnetic field is enhanced in gas gaps, while it is diminished to nearly zero in regions of high gas density.
Assuming that magnetic field enhancements are associated with stronger MRI and increased turbulence, this nonideal MHD effect of gas can be modeled by setting the viscosity inversely proportional to gas accumulation, which implies $\phi_{\rm g}=-1$.
However, note that the magnetically dead zone and the resulting layered accretion in a protoplanetary disk can also be modeled with an $\alpha$-parameter that is inversely proportional to the gas surface density \citep{Gammie1996, Kadam19}.
Fig. \ref{fig:Pl_MHD} shows the evolution of the vortex cascade when we assumed the effects of gas MHD on disk viscosity.
The general evolution of the vortex cascade is similar to model \simname{Pl\_STD}, with some notable differences.
At the snow region, a band or ring of dust-gas accumulation is formed instead of vortices, which migrates inward at a much slower rate. 
These bands are marked in red in the second panel of Fig. \ref{fig:Pl_MHD}, between the snow region and the inner boundary.
The number of vortices detected in this case corresponds to the region outside of 2.5 au.
The temperature plateau associated with the snow region forms a bump in gas density profile, while the associated pressure maximum in the gas also attracts dust. 
Since the viscosity is now inversely proportional to both dust and gas accumulation, this band-like structure evolves much slower at the local viscous timescale.
The process is similar to formation of dusty rings at the inner edge of a dead zone \citep{Kadam+22}, where in this case, the snow region provides the initial perturbation. 
However, the bands in the case of \simname{Pl\_MHD} are not fully axisymmetric and show local, dynamical clumping of both dust and gas.
These azimuthal asymmetries in the dust-gas bands are most likely related to the dependence of viscosity on dust and dust backreaction.

The third panel of Fig. \ref{fig:Pl_MHD} shows the fragmentation size of dust particles, which can exceed 20 meters at times. 
This can be explained by large gas density in the dust-gas bands, in combination with low viscosity and resulting decrease in sound speed from the lack of viscous heating (see Eq. \ref{eq:afrag}).
Although it is difficult to claim that such sizes are indeed possible with collisional dust evolution, this gives us an idea of favorable conditions for dust growth inside the vortices and, in this case, dust-gas bands. 
The mass available for SI shows a peculiar sawtooth pattern, where the peaks correspond to the occurrence of dust-gas bands inside of the snow region.
These bands are conducive for SI and the abrupt decline in $M_{\rm SI}$ is due to their exit through the inner boundary.

\begin{figure}
\centering
\includegraphics[width=0.45\textwidth]{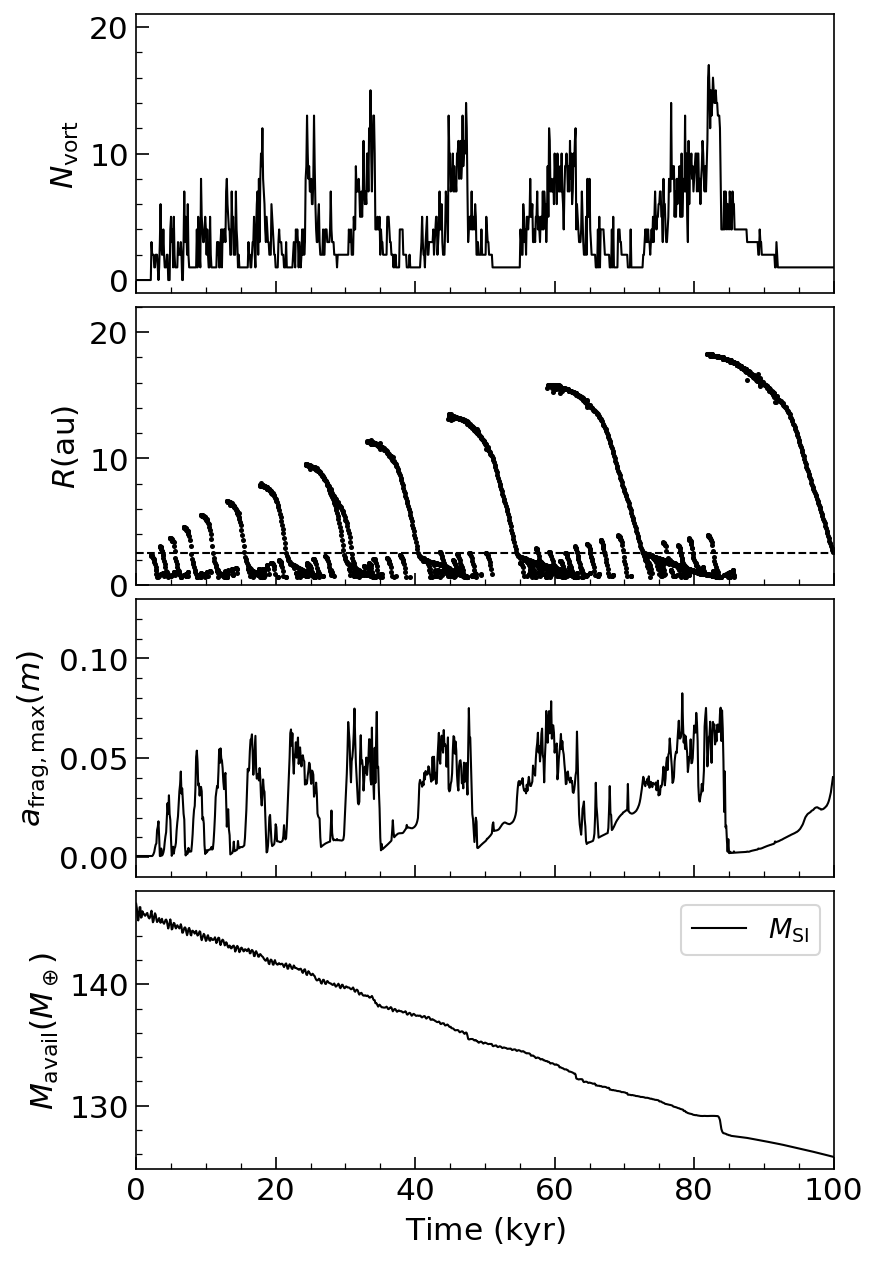} \\ 
\caption{Same as Fig. \ref{fig:Pl_STD}, but for model \simname{Pl\_Z}.}
\label{fig:Pl_Z}
\end{figure}

\subsubsection{Effects of metallicity and drawback of constant Stokes approximation}
\label{sec:z}

With the help of \simname{Pl\_Z}, we explored the vortex cascade in a high metallicity environment, where the initial dust-to-gas ratio was increased to $\zeta_{d2g, 0}=0.05$.  
The frequency of giant exoplanets is strongly correlated with the metallicity of the host star, indicating that such an environment promotes planetesimal formation \citep{Gonzalez97,Fischer-Valenti05}. 
As seen in Fig. \ref{fig:Pl_Z}, the overall behavior of the vortex cascade, in terms of their number and spacetime evolution is similar to previous models.
However, in this case, the initial disk configuration itself is streaming unstable, even before the simulation begins. 
This is seen in the last panel of Fig. \ref{fig:Pl_Z}, where the mass available for SI does not start with zero, but a large value, which decreases monotonically with time. 
At the given Stokes number of 0.01, the criterion for SI (Eq. \ref{eq:SI}) is satisfied over a large extent of the disk.
We repeated the simulation with a more moderate value of $\zeta_{d2g, 0}=0.02$, and the results with respect to SI were similar.
Three-dimensional dust-gas coupled simulations show that particle clumping with SI depends strongly on the metallicity \citep{Johansen+09}.
However, the onset of SI is not so straightforward as there are some indications that SI is unlikely to form planetesimals from millimeter grains even inside axisymmetric pressure bumps \citep{Carrera-Simon22}. 
Thus, a protoplanetary disk that is prone to SI in its entirety is not expected.

This highlights one of the limitations of our study: the approximation of a constant Stokes number for the dust component. 
Consider a dust particle at some radial distance in a canonical protoplanetary disk that moves radially with a constant Stokes number.
If this particle drifts inward or outward, it must adapt its aerodynamic cross section according to the gas surface density, which can vary by orders of magnitude over this range.
Thus, in addition to lacking collisional growth and fragmentation as well as variations in grain size distribution, a constant Stokes approximation oversimplifies dust-gas coupling.   
The approximation of constant Stokes number may not be entirely accurate for predicting behavior of the dust component over a large, radially extended disk.
Additionally, the criterion used for SI has similar drawbacks, as it is derived from local shearing-box simulations of inner disk region, with limited gas physics and particles having a constant Stokes number \citep{Yang+17}.

\subsection{Constant dust size suggesting delayed onset}
\label{sec:constsize}

\begin{figure}
\centering
\includegraphics[width=0.44\textwidth]{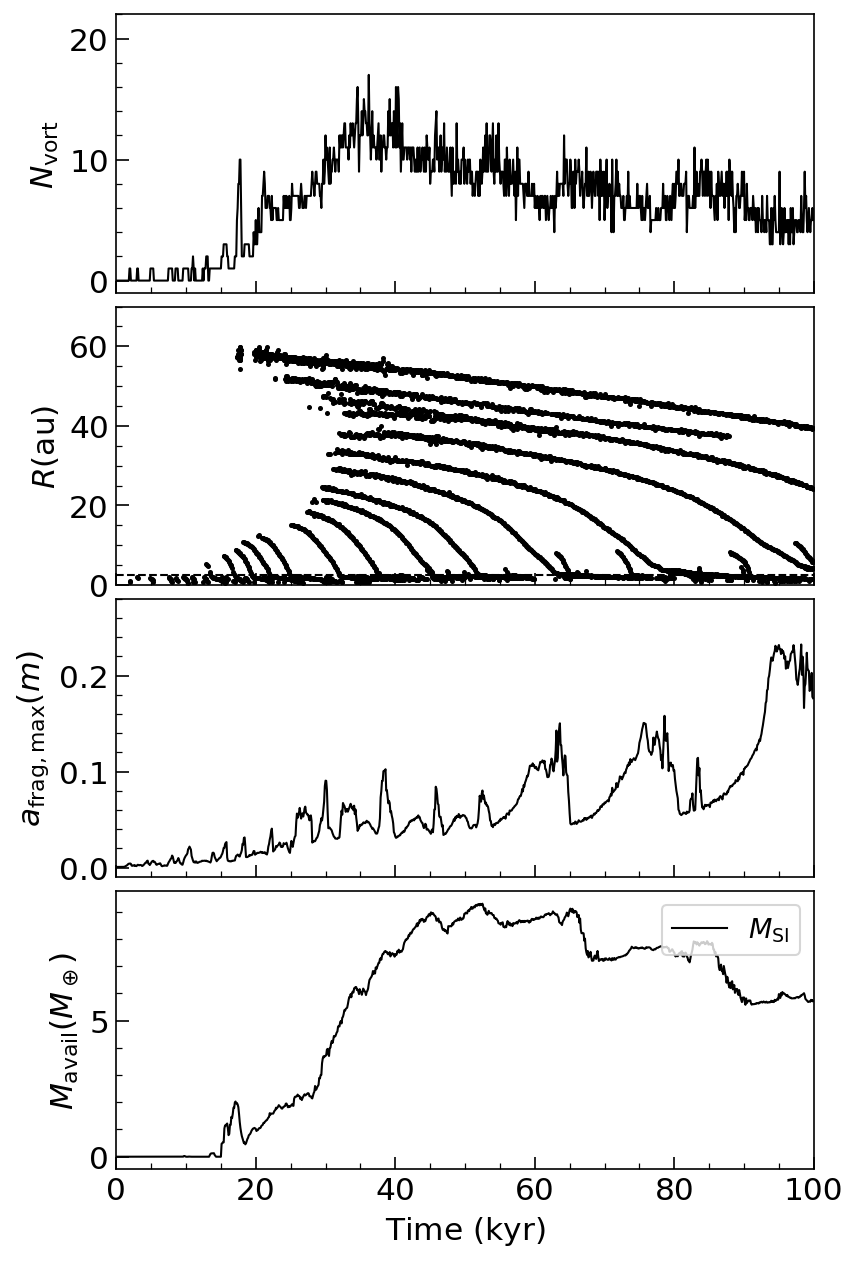} \\ 
\caption{Temporal evolution of the aggregate properties of the vortices formed in model \simname{Pl\_1mm}, showing the number of vortices, their position, the mass available for SI, and the maximum Stokes number.}
\label{fig:Pl_1mm}
\end{figure}

In order to confirm the robustness of vortex cascade with dust-dependent viscosity, we conducted additional simulations with constant dust particle size, instead of constant Stokes number (see Table \ref{table:sims}).
Fig. \ref{fig:Pl_1mm} shows the results for model \simname{Pl\_1mm}, where the vortex cascade clearly develops. 
A mm-sized particle corresponds to Stokes number of approximately 0.03 and 0.07 at the inner and outer edge, respectively, of an unperturbed disk.
The initial low number of vortices increases up to about 15 and then slowly decreases. 
However, as seen in the second panel, there is a significant difference where the vortices form, in comparison with previous constant Stokes number simulations.
The snow region initially gives rise to vortices in its vicinity, as expected, but additionally, vortices begin to form in the outer disk, giving rise to new generations that develop in an outside-in fashion. 
As we saw in the case of model \simname{Pl\_DG}, a higher Stokes number is very conducive of vortex formation.
In the outer regions, a dust particle of constant size has a much larger Stokes number because of the lower gas density in its neighborhood.
We hypothesize that small gradients in dust or gas density near the disk cutoff radius of 50 au are sufficient to trigger vortex cascade, in the case of model \simname{Pl\_1mm} with a constant particle size. 
Within these vortices, the fragmentation size supports dust growth over 20 cm, while the conditions are also suitable for SI and planetesimal formation.

\begin{figure}
\centering
\includegraphics[width=0.45\textwidth]{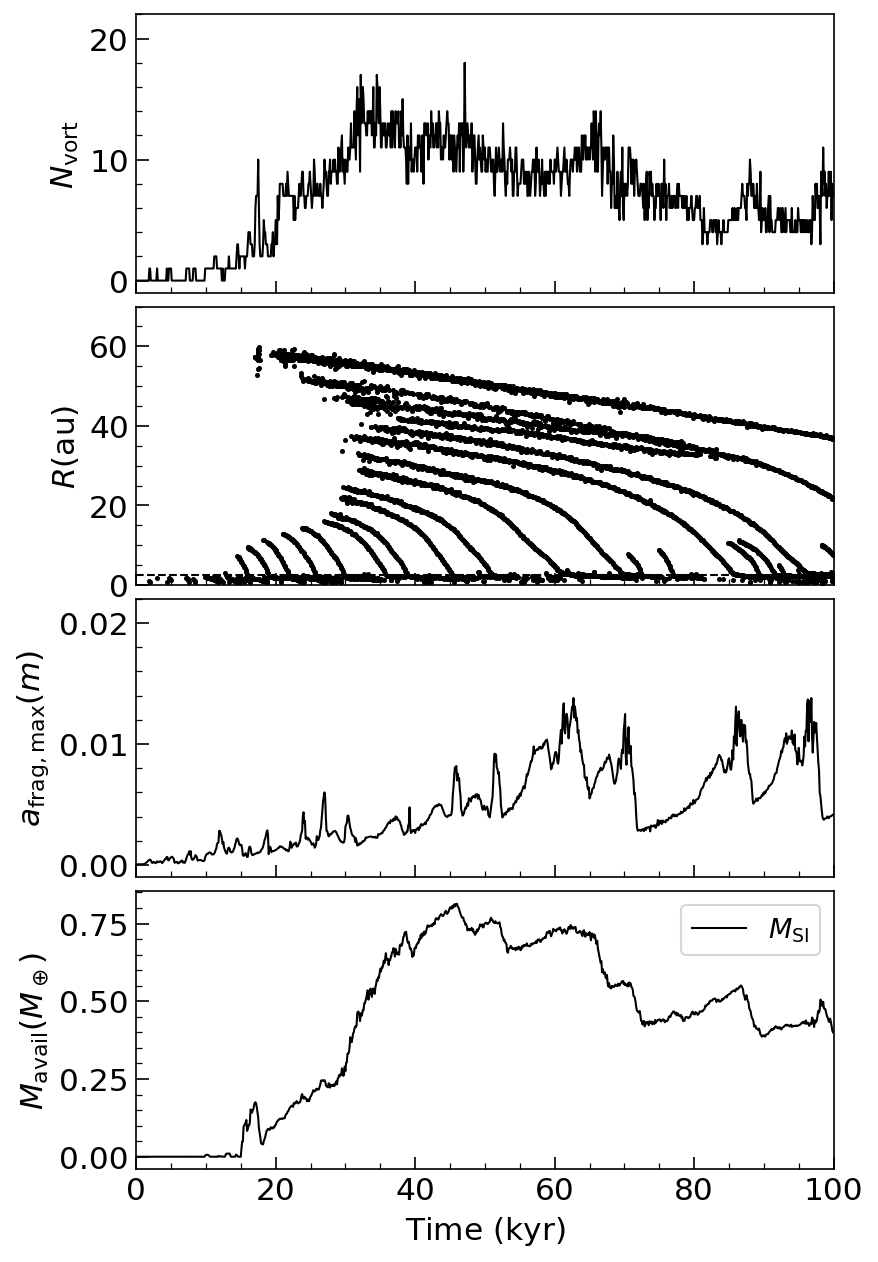} \\ 
\caption{Temporal evolution of the aggregate properties of the vortices formed in model \simname{Pl\_0.1mm\_LM}, showing the number of vortices, their position, the mass available for SI, and the maximum Stokes number.}
\label{fig:Pl_0.1mm}
\end{figure}

Millimeter-sized dust is relatively large for it to adsorb the charged particles efficiently, as most of the area for this process is provided by the small dust \citep{Balduin+23}.
In order to find out if vortices are produced with a smaller sized dust component, we used model \simname{Pl\_0.1mm}.
Here the constant dust size of 0.1 mm is more suitable for an efficient charge adsorption. However, in this case, no vortex cascade developed.
This is a conundrum for planetesimal formation via self-sustaining vortex cascade: smaller dust sizes favor dust charging, but the associated low Stokes number inhibits vortex formation, and vice versa.
However, it is possible to have a larger Stokes number for a given sized particle in a disk with lower gas density.
Hence, we modeled a lower-mass disk with model \simname{Pl\_0.1mm\_LM}, where the disk mass is 0.001 $M_{\odot}$, which is ten times smaller than in the previous models.
As shown in Fig. \ref{fig:Pl_0.1mm}, a vortex cascade indeed develops, which is remarkably similar to model \simname{Pl\_1mm}.
Due to the lower dust content in the disk, the fragmentation size and mass available for SI are about ten times smaller as compared to \simname{Pl\_1mm}.

The simulations with constant particle size imply that the vortex cascade may not be active in a protoplanetary disk in its infancy, when the gas surface density is large.
At this time, the Stokes number for a given dust species is small, which suppresses vortex formation.
As the disk evolves and its gas mass decreases, certain perturbations in the disk, in this case the temperature plateau at the water snow region, may trigger a self-sustaining vortex cascade, seeding the entire disk with planetesimals or planetary embryos in a relatively short time.
Such a delayed onset of planetesimal formation may provide us the missing link between protoplanetary disks and planetary systems.

\section{Conclusions}
\label{sec:con}

In this paper, we conducted numerical hydrodynamical experiments to investigate the possibility of vortex formation at the water snow region in a protoplanetary disk. 
The dust-dependence of the turbulent viscosity model that we implemented is physically motivated by the fact that dust particles can adsorb charges and thus reduce the local gas-magnetic field coupling necessary for the action of the MRI. 
A snow region extending about 1 au is thought to form instead of a snow line, because of the opacity changes and gradual sublimation of water at its sublimation temperature of 160 K.  
The initial conditions for the simulations are similar to an MMSN disk with a temperature plateau at the water snow region, while the gas structure was constructed to obtain a steady state accretion disk of size 100 au.
We show that self-sustaining vortices originate at the snow line in a majority of the scenarios, with a subsequent vortex cascade extending across a large portion of the disk.
The conditions that inhabited vortex formation via this mechanism of dust-dependent viscosity are large turbulent $\alpha$ (\simname{Pl\_ALPHA}), or small-sized dust that is strongly coupled with the gas (\simname{Pl\_SM} and \simname{Pl\_0.1mm}).
We conducted simulations with both a constant Stokes number and a constant particle size and demonstrate that the phenomenon of self-sustaining vortices is robust.
The models show vortex cascade for the entire 0.1 Myr of the disk evolution.

The vortices allow for large fragmentation sizes; hence, the collisional growth of dust inside the vortices can possibly produce meter-sized boulders.
In combination with the fact that the interior of vortices typically satisfy conditions for SI and their secular stability over hundreds of orbits, they would form exceptionally suitable conditions for the formation of planetesimals.
If the Stokes number is sufficiently large, the local dust density exceeds Roche density, suggesting that planetesimal formation may result in gravitationally bound planetary embryos at the location of the original vortex.
Thus, this process can seed the entire disk with planetary cores in a relatively short time.

Additionally, we show that small dust of constant size, which does not produce vortices in an MMSN disk, can produce a vortex cascade in a less massive disk (\simname{Pl\_0.1mm\_LM}).
For the vortex cascade to occur, a certain degree of decoupling between dust and gas is necessary, i.e., a sufficiently large Stokes number, which is possible for a given size of dust in a less massive disk. 
This brings forth the idea of a delayed onset for the formation of vortices in protoplanetary disks, as follows. 
At early times at the Class 0 stage, the gas surface density of the disk is so large that the dust particles have a small Stokes number, and no vortices can be produced.
With time, the dust grows and the lower gas surface density results in sufficiently large Stokes number particles.
Once this decoupling is achieved, the water snow region can suddenly trigger vortex formation cascade via dust-dependent viscosity. 
Planetesimal formation and subsequent planetary core formation follows rapidly.
This scenario can result in configurations similar to those used for the initial conditions of population synthesis methods, consisting of a gaseous disk containing several planetary core \citep{Mordasini18}.

Here, we list some of the shortcomings of our approach. 
As discussed in Section \ref{sec:z}, the constant Stokes number approximation for the dust component has its caveats; similarly, constant-sized particles also do not faithfully portray the behavior of the dust.
In order to simulate realistic dust behavior in protoplanetary disks, modeling the collisional evolution of the dust is necessary.
Although the formation of the vortex cascade appears robust for a variety of scenarios, it needs to be verified with a dust growth model.
It is possible that the fragmentation velocity is lower for unequal-mass collisions, which would delay the growth of dust grains \citep{Hasegawa+21}.
The second drawback of our findings is the assumption of dust charging and its efficiency for dust particles of larger size. 
Most of the contribution to the charge adsorption comes from small dust grains, as they offer a much larger surface area compared to a few larger dust grains \citep{Balduin+23}. 
However, the vortex cascade requires sufficiently large Stokes numbers and dust-gas decoupling.
Additionally, when dust particles are charged, their growth may be hindered by a charging barrier \citep{Okuzumi09}.
The exact values of the exponents, $\phi_{\rm g}$ and $\phi_{\rm d}$, should also be informed by studies that consider the microphysics of dust charging, which is inevitably linked to the disk's thermo-chemistry \citep{Balduin+23}. 
There are additional limitations to our model, for example the stability of vortices in 3D and effects related to the magnetic field, including nonideal MHD effects or disk winds.
We plan to address some of these issues in the near future.
In the meantime, the phenomenon of delayed-onset vortex cascade at the water snow region paints a tantalizing picture of planet formation, wherein, at the appropriate time, the disk rapidly fills with a number of vortices that not only aid planetesimal formation but also seed the disk with planetary embryos.

\begin{acknowledgements}
We thank the anonymous referee for constructive feedback, which improved the quality of this article. We thank Peter Woitke and Eduard Vorobyov for insightful discussions. K.K. acknowledges the use of IWF's leo10 cluster and VBC's CLIP cluster for GPU resources.
We acknowledge support from the COST Action CA22133 PLANETS. This research has made use of the Astrophysics Data System, funded by NASA under Cooperative Agreement 80NSSC21M00561.

\end{acknowledgements}

\bibliographystyle{aa}
\bibliography{references}

\end{document}